\documentclass[preprint,showpacs,aps,amsmath,showkeys]{revtex4}

\usepackage[cp1251]{inputenc}
\usepackage[russian]{babel}

\usepackage{xcolor}
\usepackage{latexsym,amsmath,amssymb,amsbsy,graphicx}
\usepackage{booktabs}

\makeatother

\begin{document}

    \title{Анализ различий каталогов межпланетных корональных выбросов массы и создание объединенного каталога}

    \author{А.\,О. \surname{Ширяев}$^\dagger$,$^\dagger$$^\dagger$}
    \email{anton.o.shiryaev@gmail.com}
    \author{К.\,Б. \surname{Капорцева}$^\dagger$,$^\dagger$$^\dagger$$^\dagger$}
    \affiliation{$^\dagger$
        Московский государственный университет имени М.\,В.~Ломоносова, Научно-исследовательский институт ядерной физики им. Д.В. Скобельцына. Россия, 119991, Москва, Ленинские горы, д.~1, стр.~2.
    }
    \affiliation{$^\dagger$$^\dagger$
        Брянский государственный технический университет, Факультет информационных технологий. Россия, 241035, Брянск, бул. 50 лет Октября, 7
    }
    \affiliation{$^\dagger$$^\dagger$$^\dagger$
        Московский государственный университет имени М.\,В.~Ломоносова, Физический факультет, кафедра физики космоса. Россия, 119991, Москва, Ленинские горы, д.~1, стр.~2.}

    \date{\today}

    \preprint{Ученые записки физического факультета МГУ}

    \begin{abstract}
        Для определения корональных выбросов массы (КВМ), распространяющихся в гелиосфере, используются данные о магнитных и кинетических параметрах плазмы солнечного ветра.
        Существуют различные каталоги межпланетных КВМ (МКВМ), но в связи с различиями в реализуемых в них критериях идентификации МКВМ определяемые события и их параметры могут значительно расходиться.
        В настоящей работе рассмотрены каталоги МКВМ Ричардсона и Кейн и CCMC CME Scoreboard, а также каталог крупномасштабных событий солнечного ветра ИКИ РАН, и предложен алгоритм объединения каталогов для создания более полной базы МКВМ.
        Построен объединённый каталог за 2010--2022 годы.
        Произведено дополнение объединённого каталога данными OMNI\@.
        Анализ полученного объединённого каталога показал высокую точность объединения совпадающих между каталогами событий и тенденцию событий, определённых во всех рассмотренных каталогах, к большей геоэффективности, скорости и продолжительности.
        Список опубликован на сайте центра космической погоды НИИЯФ МГУ: \url{https://swx.sinp.msu.ru/tools/icme_list.php}
        \\УДК: 523.62-726

    \end{abstract}


     \maketitle



    \section{Введение}\label{intro}
    Корональные выбросы массы (КВМ) - мощные события солнечной активности, наблюдаемые с помощью коронографа в видимом свете.
    Коронографы позволяют наблюдать за распространением КВМ на малых расстояниях от Солнца.
    Поле зрения коронографов LASCO на космическом аппарате (КА) SOHO составляет от $1.1 \ R_{solar}$ (радиусов солнца) до $3 \ R_{solar}$, от $2 \ R_{solar}$ до $6 \ R_{solar}$ и от $3.7 \ R_{solar}$ до $30 \ R_{solar}$ для LASCO C1, C2 и C3 соответственно~\cite{Wang1998}.
    Чтобы наблюдать КВМ на больших расстояниях, можно использовать такие камеры, как Heliospheric Imager на КА STEREO (поля зрения $12-84 \ R_{solar}$ и $66-318 \ R_{solar}$) или Solar Orbiter (переменное поле зрения, обусловленное эллиптической орбитой).
    Однако эти КА находятся в постоянном движении, что ограничивает возможность непрерывного мониторинга КВМ, необходимого для прогноза космической погоды.
    Помимо прямых наблюдений КВМ в видимом свете, существуют косвенные наблюдения.
    Так как плазма КВМ часто бывает плотнее плазмы фонового солнечного ветра, есть возможность определять местоположение КВМ в гелиосфере с помощью радионаблюдений~\cite{Chashey2018} и наблюдений космических лучей~\cite{Osetrova2019}.
    Еще одним способом исследования КВМ является анализ параметров плазмы солнечного ветра in-situ по кинетическим и магнитным параметрам, например, на приборах на КА Wind, ACE, DSCOVR в точке L1.
    Из-за различных способов наблюдения КВМ установление взаимно однозначного соответствия между КВМ, наблюдаемым в коронографе в видимом свете, и КВМ, определенном косвенными методами или измерениями in-situ в межпланетной среде, затруднено, поэтому последние получили название межпланетных КВМ (МКВМ).
    КВМ и МКВМ представляют собой одно и то же явление, но наблюдаются разными методами, описываются различными характерными параметрами и фиксируются в различных каталогах.

    КВМ, наблюдаемые в межпланетном пространстве, получили название межпланетных КВМ (МКВМ).
    Как один из основных носителей сильного межпланетного магнитного поля (ММП), МКВМ входят в число главных источников сильных геомагнитных возмущений~\cite{Chi2016}.
    При взаимодействии с магнитосферой Земли МКВМ вызывают магнитные бури, которые, в свою очередь, могут спровоцировать нарушение радиосвязи и навигации, наведенные токи в ЛЭП и трубопроводах, смещение радиационных поясов Земли на более низкие орбиты\cite{Tsurutani2014,Feynman2000}.
    Поэтому прогнозирование прибытия МКВМ является важной задачей.
    Моделированием распространения КВМ в гелиосфере занимаются научные группы по всему миру.
    Для для адаптации и верификации таких моделей используются данные из различных каталогов зарегистрированных у орбиты Земли МКВМ~\cite{Universe2022}.

    Структура и плазменные параметры МКВМ определяются его происхождением и условиями в межпланетной среде.
    Источниками МКВМ являются эруптивные события на Солнце, такие как выбросы протуберанцев и магнитных петель.
    МКВМ можно рассматривать как состоящий из основной части и, в зависимости от условий, области сжатия и/или ударной волны~\cite{Zurbuchen2006,Yermolaev2009}.
    Основная часть (тело) МКВМ относится к одному из двух типов - выброс массы (эджекта) или магнитное облако.
    Когда скорость МКВМ по отношению к солнечному ветру превышает скорость быстрой МГД-волны (альвеновскую скорость), на переднем фронте МКВМ возникает ударная волна~\cite{Kilpua2017}.
    Так как во время распространения в гелиосфере тело МКВМ плавно расширяется, область между телом МКВМ и ударной волной подвергается сжатию и характеризуется большими плотностью и температурой в сравнении с телом МКВМ.
    Эта область называется областью сжатия (sheath)~\cite{Kilpua2017}.
    В случае отсутствия ударной волны, если скорость КВМ ниже скорости фонового ветра, вместо области сжатия, может возникнуть область разряжения~\cite{Yermolaev2009}.

    Разные области МКВМ отличаются по плазменным характеристикам, что позволяет выделять МКВМ среди фонового ветра.
    Например, в работе~\cite{Wu2011} выделяют следующие критерии:
    \begin{itemize}
        \item повышенная напряжённость магнитного поля;
        \item плавно меняющееся направление магнитного поля;
        \item сравнительно низкая температура протонов;
        \item пониженное значение плазменного параметра $\beta$ для протонов;
        \item двунаправленные потоки тепловых электронов солнечного ветра и низкоэнергичных протонов;
        \item характерные особенности ионного состава;
        \item возникновение одно- или двухступенчатых эффектов Форбуша.
    \end{itemize}
    Применяются различные критерии определения МКВМ и связанных с ними структур по наблюдениям in situ~\cite{Chi2016}.
    Каталоги МКВМ используются для валидации моделей прогноза прибытия МКВМ~\cite{Universe2022}.
    Так как методологии составления и представленные события в различных каталогах не совпадают, использование целостного объединённого каталога МКВМ может позволить осуществить валидацию моделей применительно к более широкому диапазону явлений.
    В настоящей работе рассматриваются три каталога: каталог Richardson \& Cane~\cite{RC2003}, каталог NASA Community Coordinated Modeling Center CME Scoreboard (далее CCMC CME Scoreboard)~\cite{Riley2018} и каталог крупномасштабных явлений солнечного ветра ИКИ РАН~\cite{Yermolaev2009}.
    Использование этих трех каталогов было опробовано для валидации системы прогноза ЦДОКМ НИИЯФ МГУ в работе~\cite{Universe2022} и показало хорошие результаты.

    Каталог Richardson  \& Cane (\url{https://izw1.caltech.edu/ACE/ASC/DATA/level3/icmetable2.html}) содержит данные с 1996 года.
    В каталоге содержится информация о времени регистрации прибытия ударной волны ($T_{shock}$) и интервале регистрации тела МКВМ ($T_{start}$ и $T_{end}$), характеристика тела МКВМ как выброса материала или магнитного облака, средняя скорость МКВМ, среднее значение магнитного поля в МКВМ, максимальная скорость солнечного ветра и минимум значения индекса $D_{st}$ за период от начала до конца МКВМ.
    Применительно к части событий указано время наблюдения источника МКВМ ($T_{source}$).
    Поиск источников МКВМ производится с помощью коронографа LASCO КА SOHO или коронографов КА STEREO авторами каталога, а так же с использованием информации из статьи~\cite{Wood2017}, базы данных DONKI (\url{https://kauai.ccmc.gsfc.nasa.gov/DONKI/}) и др.
    Для выделения МКВМ в потоке солнечного ветра каталог Richardson \& Cane использует такие сигнатуры, как~\cite{RC2003}:
    \begin{itemize}
        \item изменение отношения наблюдаемой температуры протонов к ожидаемой температуре солнечного ветра;
        \item уменьшенные флуктуации и организованность магнитного поля;
        \item присутствие межпланетных ударных волн;
        \item присутствие двунаправленных потоков электронов (BDE).
    \end{itemize}

    В каталоге CCMC CME Scoreboard (\url{https://kauai.ccmc.gsfc.nasa.gov/CMEscoreboard/}) присутствуют данные с 2013 года.
    Основной задачей CCMC (общественно-координируемого центра моделирования) CME Scoreboard является сравнение методов прогнозирования прибытия МКВМ с использованием таких моделей, как drag based model~\cite{Dumbovic2021} и WSA-ENLIL~\cite{Riley2018}.
    В целях верификации каталог предоставляет информацию о наблюдении успешно спрогнозированных МКВМ.
    В каталоге имеются данные о времени наблюдения источника события ($T_{source}$) и времени регистрации прибытия межпланетной ударной волны ($T_{shock}$).
    События определяются экспертами группы CCMC или берутся из базы DONKI\@.

    Каталог крупномасштабных событий солнечного ветра ИКИ РАН (\url{http://www.iki.rssi.ru/omni/catalog/}) содержит информацию с 1976 года.
    Авторы каталога предлагают классификацию типов солнечного ветра~\cite{Yermolaev2009}, которая включает межпланетные ударные волны (IS), область сжатия (SHEATH) и тело МКВМ (EJE и MC).
    Каталог содержит данные о типе солнечного ветра и продолжительности наблюдения для каждого события.
    Для определения типа солнечного ветра в каталоге используются следующие данные из базы OMNI:
    \begin{itemize}
        \item параметры магнитного поля Земли (GSE и GSM);
        \item плотность, скорость и температура плазмы;
        \item характеристики потоков заряженных частиц;
        \item количество солнечных пятен;
        \item индексы геомагнитной активности $K_p$, $D_{st}$ и $C9$.
    \end{itemize}
    Эти сведения дополняются вычисленными параметрами солнечного ветра, исходя из которых определяется тип потока~\cite{Yermolaev2009}.


    \section{Алгоритм объединения каталогов}
    \label{Sec1}
    Объединение событий из каталогов происходит на основе проверки совпадения временных параметров МКВМ, указанных в каталогах - времени наблюдения источника $T_{source}$, времени регистрации прибытия ударной волны $T_{shock}$ и времени начала регистрации тела МКВМ $T_{start}$.
    Порядок объединения каталогов не имеет значения с точки зрения результата, но алгоритмически осуществляется последовательно: сначала производится объединение каталогов Richardson \& Cane и CCMC CME Scoreboard, затем результат их объединения дополняется событиями из каталога крупномасштабных событий солнечного ветра ИКИ РАН.
    Если события из разных каталогов удовлетворяют критериям объединения (описанным ниже), то они считаются одним событием, за параметры которого принимаются параметры из приоритетного каталога.
    Установлен следующий приоритет каталогов: Richardson \& Cane, CCMC CME Scoreboard, ИКИ РАН.
    Если события из разных каталогов не были объединены, они попадают в итоговый каталог как отдельные события.
    За рассматриваемый период времени (2010 - 2022 гг) Richardson \& Cane содержит 224 события, CCMC CME Scoreboard 192 события и каталог крупномасштабных событий солнечного ветра ИКИ РАН - 467 событий (см.\ раздел~\ref{ras}).

    \subsection{Слияние каталогов Richardson \& Cane и CCMC CME Scoreboard}
    \begin{figure}[htbp!]
        \centerline{\includegraphics{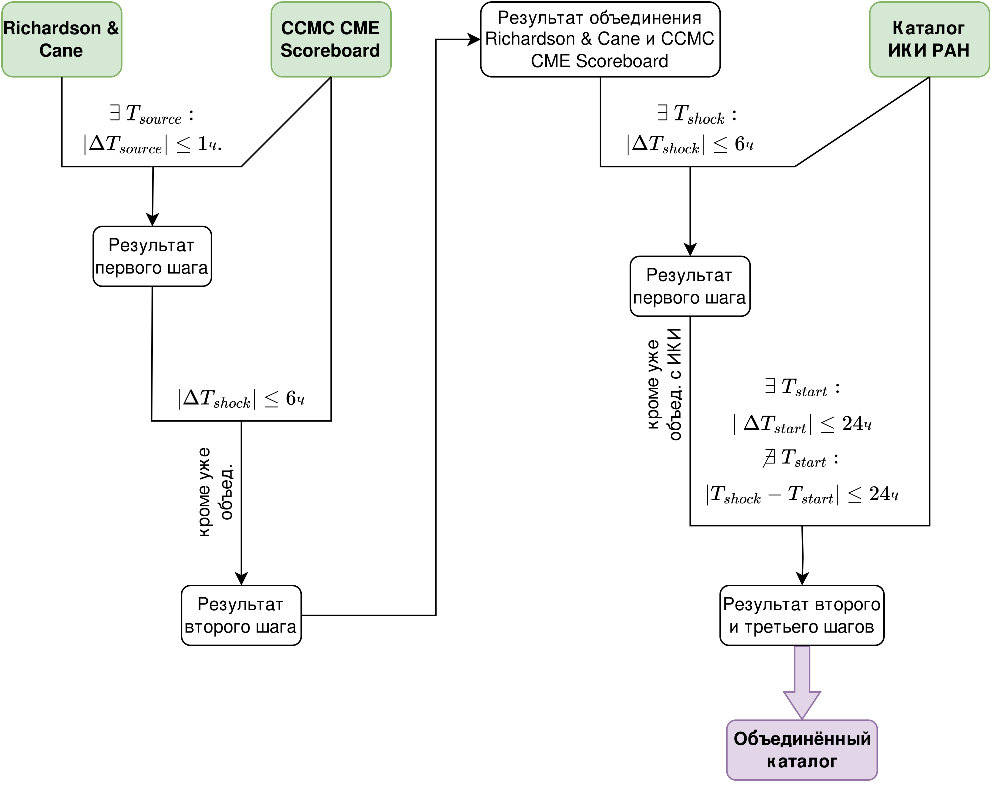}} 
        \caption{Алгоритмы объединения каталогов. Слева: объединение каталога Richardson \& Cane с CCMC CME Scoreboard. Справа: объединение каталогов Richardson \& Cane и CCMC CME Scoreboard с каталогом крупномасштабных событий солнечного ветра ИКИ РАН} \label{fig:rc_ccmc_scheme}
    \end{figure}

    Первым этапом объединения каталогов Richardson \& Cane и CCMC CME Scoreboard (рис.~\ref{fig:rc_ccmc_scheme}) является объединение по $T_{source}$, так как в большинстве событий обоих каталогов, содержащих информацию о времени наблюдения источника, оно определяется по данным коронографа LASCO\@.
    Допуск при объединении по времени наблюдения источника определяется условием $\left| \Delta T_{source} \right| \leq 1  \ \text{ч}$.
    CCMC CME Scoreboard во всех случаях предоставляет информацию о времени наблюдения источника события, в то время как Richardson \& Cane содержит информацию о времени наблюдения источника только для части событий.
    В каталогах Richardson \& Cane и CCMC CME Scoreboard указывается время регистрации прибытия ударной волны ($T_{shock}$) для всех событий.
    Поэтому вторым этапом объединения каталогов Richardson \& Cane и CCMC CME Scoreboard, для событий, которые ещё не объединены, является объединение по времени регистрации прибытия ударной волны (рис.~\ref{fig:rc_ccmc_scheme}) c допуском $\left| \Delta T_{shock} \right| \leq 6 \ \text{ч}$.

    \subsection{Слияние объединённого каталога с каталогом крупномасштабных событий солнечного ветра ИКИ РАН}\label{ras}
    Так как каталог крупномасштабных событий солнечного ветра ИКИ РАН не выделяет МКВМ, а разделяет весь временной ряд на типы, нами было произведено выделение событий, соответствовавших проходу МКВМ.
    Рассмотрены следующие последовательности типов солнечного ветра:
    \begin{enumerate}
        \item только MC (магнитное облако) или EJE (выброс массы)
        \item IS (ударная волна), затем MC или EJE
        \item IS, SHE (область сжатия), затем MC или EJE
    \end{enumerate}
    В результате обработки каталога крупномасштабных событий солнечного ветра ИКИ РАН за рассмотренный период получено 467 событий, содержащих информацию о временных границах тела события ($T_{start}$ и $T_{end}$) и о времени регистрации прибытия ударной волны ($T_{shock}$) в случае её наличия.
    В качестве времени начала события ($T_{start}$) рассматривается время начала эджекты или магнитного облака.

    Объединение с каталогом ИКИ происходит на основе времени регистрации прибытия ударной волны и времени начала регистрации тела МКВМ (рис.~\ref{fig:rc_ccmc_scheme})
    Первым шагом объединения с каталогом крупномасштабных событий солнечного ветра ИКИ РАН является объединение по времени прибытия ударной волны, осуществляемое с допуском $\left|\Delta T_{shock} \right| \leq 6 \  \text{ч}$.
    Вторым шагом производится объединение по времени начала с допуском $\left| \Delta T_{start} \right| \leq 24 \  \text{ч}$.
    После этого остаются события из каталога CCMC CME Scoreboard, не объединённые с каталогом Richardson \& Cane.
    У таких событий указаны только время наблюдения источника события и время регистрации прибытия ударной волны.
    Для их сопоставления с событиями ИКИ, у которых указано только временные границы тела МКВМ, третьим шагом проверяются совпадения времени регистрации прибытия ударной волны согласно CCMC CME Scoreboard и времени начала регистрации тела МКВМ согласно каталогу крупномасштабных событий солнечного ветра ИКИ РАН с допуском в $\left| T_{shock} - T_{start} \right| \leq\ 24 \  \text{ч}$.


    \section{Анализ объединённого каталога}
    \label{Sec2}

    \subsection{Состав объединённого каталога}
\begin{figure}[htbp!]
        \centerline{\includegraphics{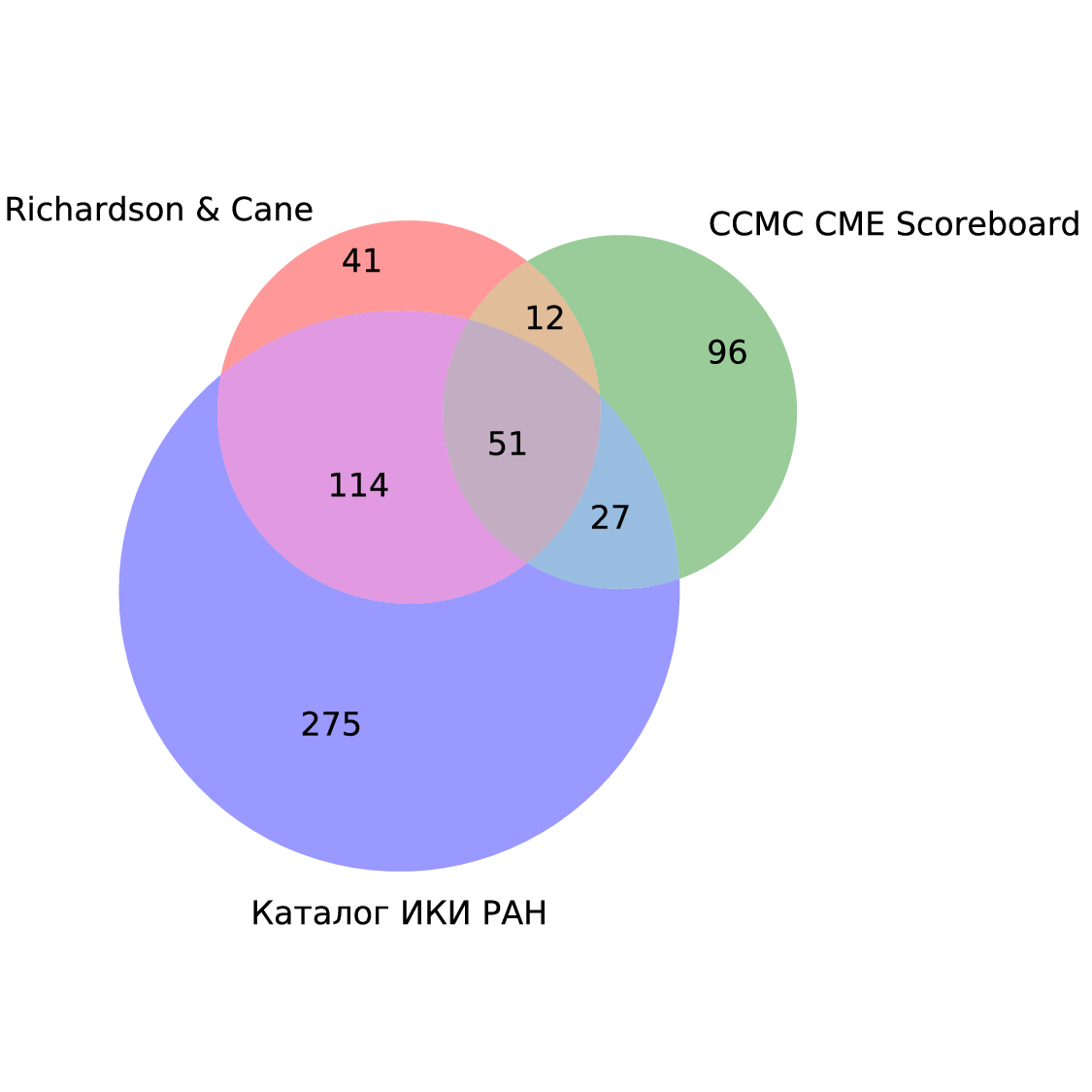}} \caption{Состав объединённого каталога. Круги соответствуют событиям из каждого каталога, пересечения кругов - объединенным событиям. Числами указано количество событий в каждом множестве} \label{fig: merged_catalogue_composition}
    \end{figure}

    Результирующий каталог содержит 622 события.
    Для анализа состава объединенного каталога было вычислено количество событий, которые были указаны только в одном, в двух и во всех трех каталогах.
    Результат представлен в виде диаграммы Венна на рис.~\ref{fig: merged_catalogue_composition}.
    Анализ состава показал, что 41 событие фигурирует только в каталоге Richardson \& Cane, 96 - только в CCMC CME Scoreboard и 275 - только в каталоге крупномасштабных событий солнечного ветра ИКИ РАН.
    18 событий -- общие для Richardson \& Cane и CCMC CME Scoreboard, 114 - для Richardson \& Cane и крупномасштабных событий солнечного ветра ИКИ РАН, и 27 - для CCMC CME Scoreboard и ИКИ РАН.
    При этом 51 событие содержится во всех трёх каталогах.
    В таблице~\ref{tab: yearly_table} рассчитано количество событий по годам для рассмотренных каталогов, объединённого каталога и для событий объединённого каталога с известным временем наблюдения источника.

    Динамика количества событий в каталоге Richardson \& Cane хорошо соответствует фазам цикла солнечной активности на рассмотренном периоде: в фазе роста и максимуме 24 цикла (2011--2015 год) определено больше событий, в период минимума солнечной активности (2010 год, 2019--2020 год) -- меньше событий.
    Аналогичная ситуация наблюдается в каталоге CCMC CME Scoreboard, но в 25 цикле (2020--2022 год) определено значительно больше событий, чем в максимуме 24 цикла.
    Это может говорить либо о большей мощности 25 цикла в сравнении с 24, либо о совершенствовании методов идентификации событий в CCMC CME Scoreboard.
    Количество событий, определённых в каталоге крупномасштабных событий солнечного ветра ИКИ РАН, не показывает такой согласованности с солнечной активностью.
    Это может быть связанно с более высокой точностью определения сравнительно более слабых событий, не обусловленных солнечными вспышками, в сравнении с другими рассмотренными каталогами.
    \begin{table}[hbtp!]
        \begin{center}
            \begin{tabular}{|c|c|c|c|c|c|}
                \hline
                \toprule
                \textbf{Год} & \textbf{Richardson \& Cane} & \textbf{CCMC CME Scoreboard} & \textbf{ИКИ РАН} & \textbf{С временем источника} & \textbf{Все} \\
                \hline
                \midrule
                2010         & 4                           & --                           & 20               & 3                             & 21           \\
                2011         & 32                          & --                           & 52               & 18                            & 62           \\
                2012         & 35                          & --                           & 53               & 21                            & 61           \\
                2013         & 25                          & 16                           & 32               & 21                            & 47           \\
                2014         & 20                          & 27                           & 28               & 28                            & 46           \\
                2015         & 29                          & 19                           & 32               & 23                            & 48           \\
                2016         & 13                          & 10                           & 35               & 10                            & 40           \\
                2017         & 9                           & 10                           & 32               & 10                            & 39           \\
                2018         & 8                           & 5                            & 39               & 5                             & 39           \\
                2019         & 7                           & 8                            & 40               & 8                             & 44           \\
                2020         & 4                           & 9                            & 26               & 11                            & 29           \\
                2021         & 10                          & 44                           & 21               & 44                            & 54           \\
                2022         & 16                          & 42                           & 23               & 42                            & 57           \\
                \hline
                Всего        & 224                         & 192                          & 467              & 250                           & 622          \\
                \bottomrule
                \hline
            \end{tabular}
        \end{center}
        \caption{Количество событий в объединённом каталоге по годам}
        \label{tab: yearly_table}
    \end{table}
    Рассмотрена средняя продолжительность МКВМ ($T_{end} - T_{start}$) для изначальных каталогов, объединённого каталога, событий объединённого каталога, определённых по нескольким каталогам, и определённым по всем трём рассмотренным каталогам.
    В связи с отсутствием данных о временных границах тела МКВМ в каталоге CCMC CME Scoreboard средняя продолжительность указанных в нём событий не могла быть вычислена.
    Средняя продолжительность события в объединённом каталоге - 19.9 часов (таб.~\ref{tab: duration_table}).
    Средняя продолжительность тела МКВМ у событий, заявленных в каталоге Richardson \& Cane, больше, чем у событий из каталога крупномасштабных событий солнечного ветра ИКИ РАН.
    Успешно объединённые события имеют в среднем большую продолжительность.
    \begin{table}[hbtp!]
        \begin{center}
            \begin{tabular}{|c|c|}
                \hline
                \toprule
                \textbf{Каталог}                       & \textbf{Средняя продолжительность} \\
                \midrule
                \hline
                Все события                            & 19.9 ч                             \\
                \hline
                Все объединённые события               & 24.7 ч                             \\
                \hline
                Объединённые по трём каталогам события & 30.7 ч                             \\
                \hline
                Richardson \& Cane                     & 24.1 ч                             \\
                \hline
                CCMC CME Scoreboard                    & нет данных                         \\
                \hline
                ИКИ РАН                                & 19.8 ч                             \\
                \hline
                \bottomrule
            \end{tabular}
        \end{center}
        \caption{Средняя продолжительность событий в зависимости от их статуса объединения}
        \label{tab: duration_table}
    \end{table}

    \subsection{Точность совпадения при объединении}
    \begin{figure}[htbp!]
        \centerline{\includegraphics{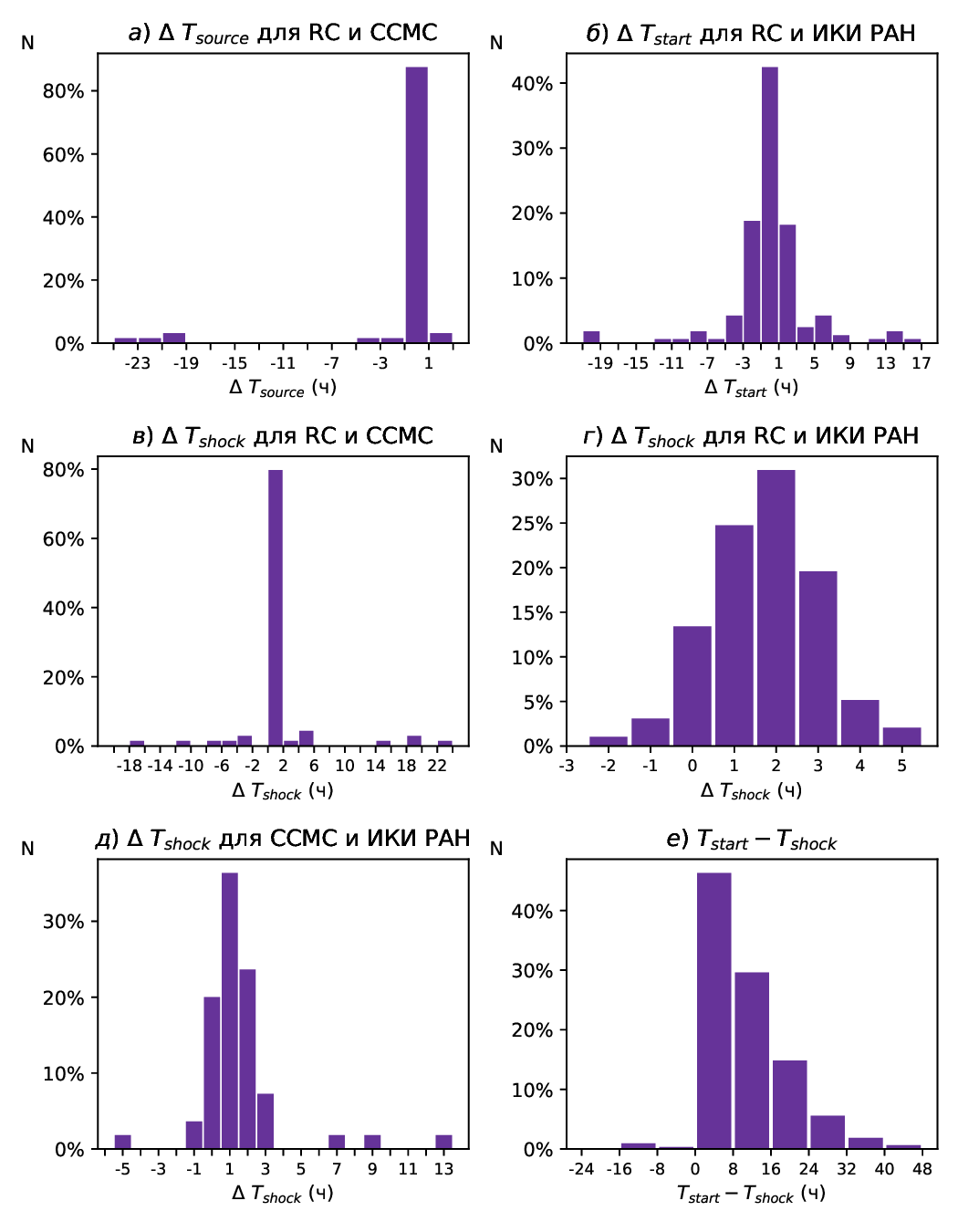}} \caption{разница а) по времени наблюдения источника между объединёнными событиями из Richardson \& Cane и CCMC CME Scoreboard б) по времени наблюдения источника между Richardson \& Cane и каталогом крупномасштабных событий солнечного ветра ИКИ РАН в-д) по времени регистрации прибытия ударной волны между каталогами Richardson \& Cane и CCMC CME Scoreboard, Richardson \& Cane и ИКИ РАН, и CCMC CME Scoreboard и ИКИ РАН соответственно е) между временем начала регистрации тела МКВМ и времени регистрации прибытия ударной волны} \label{fig: delta_plots}
    \end{figure}

    Для валидации выбранных временных допусков был рассмотрен разброс значений параметров объединяемых событий между различными каталогами.
    В большинстве случаев расхождение значений одних и тех же параметров объединённых событий согласно различным исходным каталогам значительно меньше допуска объединения (рис.~\ref{fig: delta_plots}).
    События, для которых разность значений больше установленного допуска, были объединены по другим параметрам, по которым соблюдался допуск.

    Например, на рисунке ~\ref{fig: delta_plots} \textit{а)} показана гистограмма плотности распределения разности времени определения источника согласно каталогам Richardson \& Cane и CCMC CME Scoreboard.
    Анализ показал, что более 80\% всех событий, имеющих информацию о времени источника согласно этим двум каталогам, совпадают с допуском $\pm 1 \ \text{ч}$ и $33\%$ совпадают в точности.
    События, которые совпадают в точности, предположительно, были определены исходными каталогами по одному и тому же источнику данных.
    Допуск объединения по времени источника составлял $\pm 6 \ \text{ч}$, но на рисунке ~\ref{fig: delta_plots} \textit{а)} присутствуют события, для которых отличие составляет более 19 ч.
    Таких событий 4.
    Они были объединены по времени прибытия ударной волны.
    Предположительно, эти события относят к различным корональным источникам в различных исходных каталогах.

    На рисунке ~\ref{fig: delta_plots} \textit{б)} показана гистограмма плотности распределения разности времени начала регистрации тела события для каталогов Richardson \& Cane и ИКИ РАН.
    Медиана этой величины составляет $-1.0\ \text{ч}$.
    Сравнение с плотностью распределения разности времени регистрации прибытия ударной волны между теми же каталогами, медиана которого $1.87 \ \text{ч}$, показывает более точное совпадение определения времени начала регистрации тела МКВМ в сравнении с временем регистрации прибытия ударной волны для этих каталогов.
    Время прибытия ударной волны, определённое по каталогу ИКИ РАН, опережает соответствующее время по каталогу Richardson \& Cane в среднем на $1.67 \ \text{ч}$, в то время как время начала регистрации тела МКВМ в среднем отстаёт на $0.45 \ \text{ч}$.

    На рисунке ~\ref{fig: delta_plots} \textit{в)} показана гистограмма плотности распределения разности времени регистрации прибытия ударной волны согласно каталогам Richardson \& Cane и CCMC CME Scoreboard.
    Более 75\% всех событий, объединённых таким образом, совпадают с допуском $\pm 1 \ \text{ч}$.
    Это свидетельствует о том, что наибольшая точность совпадения определения времени регистрации прибытия ударной волны наблюдается для каталогов Richardson \& Cane и CCMC CME Scoreboard.

    Значения $T_{shock}$ для событий, определённых по нескольким каталогам, совпадают в пределах $1 \ \text{ч}$ для $83\%$, $15\%$ и $29\%$ событий для Richardson \& Cane и CCMC CME Scoreboard, Richardson \& Cane и каталога ИКИ РАН и CCMC CME Scoreboard и каталога ИКИ РАН соответственно (рис.~\ref{fig: delta_plots} \textit{в-д)})

    На рисунке~\ref{fig: delta_plots} \textit{е)} показана плотность распределения длительности интервала между регистрацией ударной волны и началом регистрации тела МКВМ $T_{start}-T_{shock}$.
    Эту величину можно интерпретировать как длительность периода регистрации ударной волне и, в случае её наличия, области сжатия.
    Для $44\%$ событий это значение не превышает $8 \ \text{ч}$.
    В результате третьего шага объединения с каталогом крупномасштабных событий солнечного ветра ИКИ РАН возникают события с отрицательным значением этой величины.
    Это означает, что время прибытия ударной волны согласно CCMC CME Scoreboard наступает позже времени начала МКВМ согласно каталогу ИКИ РАН.
    Такой результат не считается ошибочным, если время прибытия ударной волны наступает раньше времени окончания регистрации тела МКВМ.
    Это условие соблюдается для 4 таких событий.
    Для одного события оно не соблюдается; объединение такого события мы считаем ошибочным и не производим несмотря на выполнение условий.

    \subsection{Сопоставление с OMNI}
    Чтобы расширить информацию о МКВМ и оценить геоэффективность событий объединённого каталога, были рассчитаны минимальное значение индекса $D_{st}$, а также максимальное и среднее значение скорости солнечного ветра по данным OMNI за период времени, соответствующий прохождению МКВМ~\cite{King2005}.
    Для событий, у которых определено время начала тела МКВМ, значения параметров вычислялись за период с $T_{start}$ до $T_{end}$.
    В случае событий, которые присутствуют только в каталоге CCMC CME Scoreboard, известно только $T_{source}$ и $T_{shock}$.
    У таких событий в качестве начала периода использовалось $T_{shock}$, а в качестве времени окончания - $T_{shock} + \left<T_{duration}\right>$, где $\left<T_{duration}\right>$ - средняя продолжительность регистрации тела МКВМ, которая в объединённом каталоге составляет 20 часов (табл.~\ref{tab: duration_table}).

\begin{figure}[htbp!]
        \centerline{\includegraphics{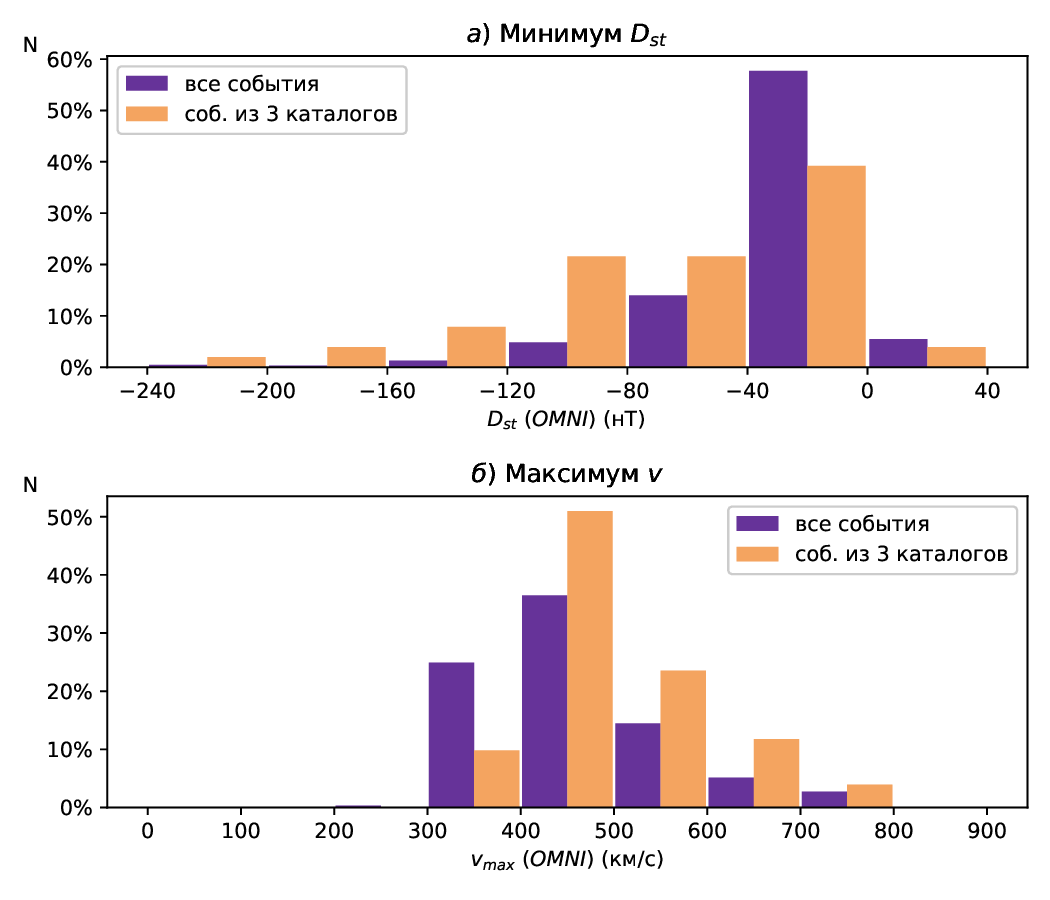}} \caption{Распределение количества событий по значениям а) минимума значения индекса $D_{st}$ б) максимума скорости солнечного ветра во время события для всех событий и событий, присутствующих во всех трёх каталогах} \label{fig: v_123_plots}
    \end{figure}

    Индекс $D_{st}$ отражает возмущение магнитного поля Земли вблизи дипольного экватора, и может рассматривается как критерий геоэффективности МКВМ~\cite{RC2010}.
    Понижение значения $D_{st}$ является характерным индикатором основной фазы магнитной бури.
    Так как тело МКВМ также может захватывать внезапное начало бури (sudden storm commencement) или фазу восстановления, которые характеризуются возрастанием $D_{st}$, рассматривается минимум значения $D_{st}$ на всём интервале.
    В момент прибытия МКВМ скорость солнечного ветра на орбите Земли можно рассматривать как непосредственную характеристику МКВМ, поэтому также была рассмотрена максимальная скорость солнечного ветра.
    Оценка скорости МКВМ важна для валидации прогнозирования прихода МКВМ.
    Для оценки геоэффективности МКВМ построены гистограммы распределения количества событий с различными значениями $D_{st}$ (рис.~\ref{fig: v_123_plots}, график \textit{a}) и максимальной скорости солнечного ветра для всех событий объединённого каталога и для событий, которые присутствуют во всех трёх рассматриваемых каталогах (рис.~\ref{fig: v_123_plots}, график \textit{б}).
    Оказывается, что $8 \%$ событий объединённого каталога вызвали возмущение $D_{st} > -75 \ \text{нТ}$.
    Для событий, присутствующих во всех трёх рассмотренных каталогах, эта величина составляет $35 \%$.
    Из этого следует, что события, которые присутствуют во всех трёх объединяемых каталогах, обладают большей геоэффективностью в сравнении с остальными событиями по рассмотренным критериям.
    Из гистограммы распределения событий с определённой скоростью следует, что доля высокоскоростных событий ($v > 400 \ \text{км/с}$) выше для событий, определённых для всех трёх каталогов ($90 \%$ против $58 \%$).
    Учитывая так же большую длительность таких событий (табл.~\ref{tab: duration_table}), можно сделать вывод о более высокой точности определения более мощных (геоэффективных, высокоскоростных и длительных) событий.

    \subsection{Доступ к данным}
    \begin{figure}[htbp!]
        \centerline{\includegraphics[scale=0.2]{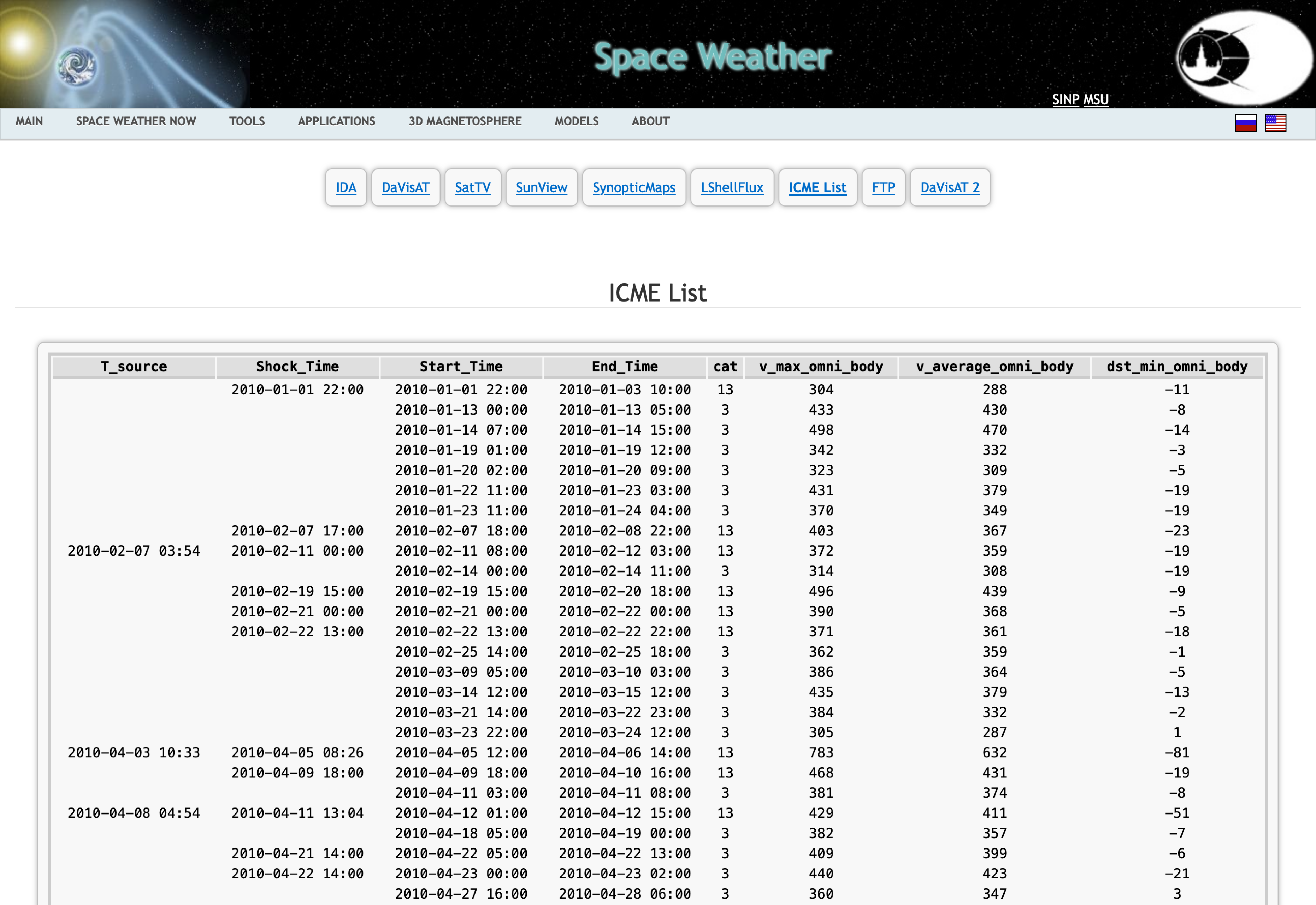}} \caption{Скриншот интерфейса объединённого каталога МКВМ, размещённого на сайте Центра космической погоды НИИЯФ МГУ по адресу \url{https://swx.sinp.msu.ru/tools/icme_list.php}} \label{fig: sinp_screenshot}
    \end{figure}
    Построенный объединённый каталог доступен на сайте центра космической погоды НИИЯФ МГУ \url{https://swx.sinp.msu.ru/tools/icme_list.php}.
    Описание колонок объединённого каталога представлено в таблице~\ref{tab:catalogue_description}.
    Скриншот интерфейса объединённого каталога представлен на рис.~\ref{fig: sinp_screenshot}.

    \begin{table}[hbtp!]
        \centering
        \begin{tabular}{|c|c|}
            \hline
            \toprule
            \textbf{Колонка}       & \textbf{Описание}                                                      \\
            \hline
            \midrule
            T\_source              & Время наблюдения источника МКВМ                                        \\
            Shock\_Time            & Время регистрации прибытия ударной волны                               \\
            Start\_Time            & Время начала регистрации тела МКВМ                                     \\
            End\_Time              & Время окончания регистрации тела МКВМ                                  \\
            cat                    & Происхождение события (см.\ описание)                                  \\
            v\_max\_omni\_body     & Максимальная $v$ солнечного ветра за интервал регистрации тела события \\
            v\_average\_omni\_body & Средняя $v$ солнечного ветра за интервал регистрации тела события      \\
            dst\_min\_omni\_body   & Минимум индекса $D_{st}$  за интервал регистрации тела события           \\
            \hline
            \bottomrule
        \end{tabular}
        \caption{Описание колонок объединённого каталога. В происхождении описания события 1 означает присутствие события в каталоге Richardson \& Cane, 2 - CCMC CME Scoreboard, 3 - каталоге ИКИ РАН. Наличие двух или трёх цифр означает присутствие события во всех каталогах, цифры которых указаны. Параметры скорости и $D_{st}$ посчитаны по базе данных OMNI.}
        \label{tab:catalogue_description}
    \end{table}

    \section{Заключение}
    Прогнозирование прибытия МКВМ является важной задачей физики космоса, имеющей приложения в таких областях, как радиосвязь и навигация.
    Для валидации моделей движения МКВМ используются каталоги МКВМ.
    Существуют различные каталоги МКВМ, которые отличаются по критериям определения событий и не совпадают между собой.
    В работе рассмотрены каталоги МКВМ Richardson \& Cane, CCMC CME Scoreboard и каталог крупномасштабных событий солнечного ветра ИКИ РАН, и представлен алгоритм их объединения.
    В основе алгоритма лежат критерии объединения событий по совпадению временных параметров с установленными в ходе исследования допусками.
    Произведено объединение каталогов и их дополнение данными о скорости солнечного ветра и величине геомагнитных возмущений, спровоцированных приходом событий из базы OMNI\@.
    Построенный объединённый каталог доступен на сайте центра космической погоды НИИЯФ МГУ \url{https://swx.sinp.msu.ru/tools/icme_list.php} и включает 622 события за 2010--2022 год.
    Треть событий присутствует в двух или более из расмотренных каталогов.

    Вычислена средняя продолжительность событий для рассмотренных каталогов, объединённого каталога, событий объединённого каталога, определённых по нескольким каталогам, и событий, определённым по всем трём рассмотренным каталогам.
    Установлено, что события, определённые во всех трёх каталогах, в среднем обладают большей продолжительностью.

    Произведена проверка точности совпадения определения параметров объединённых событий различными рассмотренными каталогами.
    Показано, что значительная часть объединённых событий совпадает между каталогами с различием, значительно меньшим установленного допуском максимального.

    Проведён анализ скорости солнечного ветра и индекса $D_{st}$ на протяжении событий объединённого каталога.
    Показано, что МКВМ, обнаруженные во всех трёх рассмотренных каталогах, обладают большей скоростью и вызывают более сильные магнитные возмущения в сравнении со средними значениями для событий объединённого каталога.

    Таким образом, события, определённые во всех трёх каталогах, в среднем имеют большую скорость, большую продолжительность и приводят к более сильным возмущениям магнитосферы Земли, чем остальные события объединённого каталога.
    Иными словами, более мощные события успешнее определяются всеми тремя каталогами.

    Рассмотрено количество представленных в рассмотренных каталогах событий в зависимости от фазы солнечной активности.
    Продемонстрировано, что различные каталоги имеют различную динамику количества определённых событий в фазе пониженной солнечной активности.

    Использование объединённого каталога для валидации модели предсказания прибытия МКВМ позволит повысить точность работы модели применительно к менее геоэффективным событиям или в периоды низкой солнечной активности.
    При этом точность предсказания более мощных событий не должна пострадать, так как анализ объединённого каталога показал высокую точность совпадения для таких событий.

    \begin{acknowledgments}
        Авторы выражают благодарность научным коллективам каталогов МКВМ Richardson \& Cane, CCMC CME Scoreboard, каталога крупномасштабных событий солнечного ветра ИКИ РАН и базы данных OMNI за предоставление доступа к данным, и профессору В.~В.\ Калегаеву за значимые советы при проведении исследования.

        Исследование проведено в НИИЯФ МГУ за счет гранта Российского научного фонда (РНФ) № 22-62-00048.
    \end{acknowledgments}



    \begin{center}
        \textbf{Analysis of Differences between ICME catalogues and Construction of a Unified Catalogue}

        \medskip\small

        \textbf{A.\,O. Shiryaev$^{1,2,a}$, K.\,B. Kaportseva$^{1,3}$}

        \smallskip

        \textit {$^1$Skobeltsyn Institute of Nuclear Physics, M.V.Lomonosov Moscow State University, Moscow 119991, Russia.}\\
        \textit {$^2$Faculty of Information Technology,
            Bryansk State Technical University, Bryansk 241035, Russia.}\\
        \textit {$^2$Department of Space Physics, Faculty of Physics, M.V.Lomonosov Moscow State University, Moscow 119991, Russia}\\
        \textit {E-mail: $^a$anton.o.shiryaev@gmail.com}

        \medskip

        Multiple magnetic and kinetic solar wind plasma parameters are used to detect coronal mass ejections (CMEs) as they travel through the heliosphere.
        There are various interplanetary CME (ICME) catalogues, but due to differences between their ICME identification criteria they can significantly vary.
        In this paper we analyze Richardson \& Cane and CMC CME Scoreboard ICME catalogues and the SRI RAS solar wind types catalogue, and propose an algorith of merging them.
        A unified catalogue is constructed for 2010 to 2022.
        The resulting catalogue is completed with data from the OMNI database.
        Analysis of the unified catalogue demonstrated high accuracy when merging events present in multiple catalogues and a tendency of events defined in all three initial catalogues to demonstrate greater duration, speed and geoeffectiveness.
        The catalogue is presented on SINP MSU's Space Weather Exchange website: \url{https://swx.sinp.msu.ru/tools/icme_list.php}.

        \smallskip

        PACS: 96.50.Ci, 94.05.Sd\\
        \textit{Keywords}: space weather, interplanetary coronal mass ejections, solar wind.\\

        \textit{Received ?? Month 2009.} 

    \end{center}

    \vspace{30pt}
    \begin{widetext}
        \bigskip

        \noindent
        \textbf{Сведения об авторах} 

        \smallskip

        \noindent 1. Ширяев Антон Олегович~--- ведущий электроник НИИЯФ МГУ, Москва, Россия, e-mail: anton.o.shiryaev@gmail.com.

        \noindent 2. Капорцева Ксения Борисовна~--- младший научный сотрудник НИИЯФ МГУ, аспирант кафедры физики космоса Физического факультета МГУ, Москва, Россия,  e-mail: kb.kaportseva@physics.msu.ru.

    \end{widetext}

\end{document}